\documentclass[11pt]{article}
\usepackage[sc]{mathpazo} 
\usepackage{fullpage}
\usepackage[utf8]{inputenc}

\usepackage{lineno}
\usepackage{color}
\usepackage{setspace}

\usepackage{titlesec}
\titleformat{\section}[block]{\Large\bfseries\filcenter}{\thesection}{1em}{}
\titleformat{\subsection}[block]{\Large\itshape\filcenter}{\thesubsection}{1em}{}
\titleformat{\subsubsection}[block]{\large\itshape}{\thesubsubsection}{1em}{}
\titleformat{\paragraph}[runin]{\itshape}{\theparagraph}{1em}{}[. ]

\usepackage{amsmath, amsthm, amssymb, amsfonts, bm}

\usepackage{graphicx}
\graphicspath{{images/}{../images/}}

\usepackage{subfig}

\usepackage{caption}
\usepackage{booktabs}    
\usepackage{multirow}
\usepackage{array}
\usepackage{rotating}

\usepackage{nameref, hyperref}

\title{
Nitrogen-induced hysteresis in grassland biodiversity: \\ a theoretical test of litter-mediated mechanisms
}

\author{Katherine Meyer$^{1}$, James Broda$^2$, Andrew Brettin$^{3}$,  Mar\'ia S\'anchez Mu\~niz$^4$, Sarah Gorman$^4$,\\ Forest Isbell$^5$, Sarah E. Hobbie$^5$, Mary Lou Zeeman$^6$, Richard McGehee$^4$}

\date{}

\begin{document}

\maketitle

\noindent 1. Department of Mathematics, Carleton College, Northfield, Minnesota 55057\\
\noindent 2. Department of Mathematics, , Washington and Lee University, Lexington, VA 24450\\
\noindent 3. Center for Atmosphere Ocean Science, Department of Mathematics, Courant Institute of Mathematical Sciences, New York University, New York, New York 10012\\
\noindent 4.  School of Mathematics, University of Minnesota, Minneapolis, Minnesota 55455\\
\noindent 5. Department of Ecology, Evolution and Behavior, University of Minnesota, Saint Paul, Minnesota 55108\\
\noindent 6.  Mathematics Department, Bowdoin College, Brunswick, Maine 04011\\

\vfill

\modulolinenumbers[2]

\newpage{}

\section*{Abstract}
\label{sec:abstract}

The global rise in anthropogenic reactive nitrogen (N) and the negative impacts of N deposition on terrestrial plant diversity are well-documented. 
The R* theory of resource competition predicts reversible decreases in plant diversity in response to N loading. 
However, empirical evidence for the reversibility of N-induced biodiversity loss is mixed. 
In a long-term N-enrichment experiment in Minnesota, a low-diversity state that emerged during N addition has persisted for decades after additions ceased. 
Hypothesized mechanisms preventing recovery of biodiversity include nutrient recycling, insufficient external seed supply, and litter inhibition of plant growth.
Here we present an ODE model that unifies these mechanisms, produces bistability at intermediate N inputs, and  qualitatively matches the observed hysteresis at Cedar Creek.
Key features of the model, including native species' growth advantage in low-N conditions and limitation by litter accumulation, generalize from Cedar Creek to North American grasslands.
Our results suggest that effective biodiversity restoration in these systems may require management beyond reducing N inputs, such as burning, grazing, haying, and seed additions. 
By coupling resource competition with an additional inter-specific inhibitory process, the model also illustrates a general mechanism for bistability and hysteresis that may occur in multiple ecosystem types. 

\newpage{}

\section*{Introduction}
\label{sec:introduction}

Biodiversity supports many ecosystem functions on which humans depend \cite{balvanera2006quantifying, cardinale2012biodiversity, hooper2005effects, isbell2011high}.  
In particular, species richness (see Box 1) in plant communities tends to increase productivity 
\cite{
flombaum2008higher, marquard2009positive} and to stabilize this productivity through time \cite{
isbell2015biodiversity, loreau2013biodiversity}. 

Biodiversity has declined markedly over the past century \cite{ceballos2015accelerated,newbold2015global,pimm2014biodiversity}, driven by human impacts including reactive nitrogen (N) inputs from increasing fertilizer application and fossil fuel combustion (\cite{galloway2004nitrogen, erisman2013consequences} and references therein).  

During the past two decades, however, dry N deposition has declined in the eastern United States, western Europe, and Japan \cite{jia2016global}.
Along with reduction in emissions from fossil-fuel combustion, modification and modernization of agricultural practices hold great potential to mitigate humankind's influence on the N cycle \cite{vitousek1997human}. In the context of reductions to reactive N inputs, the degree to which N-induced biodiversity losses are reversible may have long-term implications for ecosystem function.

Observations and experiments in grassland sites across North America \cite{avolio2014changes,harpole2007grassland, huenneke1990effects, isbell2013nutrient,simkin2016conditional, suding2005functional} and in Great Britain \cite{stevens2004impact,silvertown2006park} have established consistent patterns and mechanisms of N-induced declines in species richness.
The reversibility of N-induced biodiversity losses, however, remains unclear, based on divergent findings at Cedar Creek Ecosystem Science Reserve in Minnesota \cite{isbell2013low} and the Rothamsted Park Grass Experiment in Great Britain \cite{storkey2015grassland}. The Cedar Creek experiment began in 1982 at a late-successional grassland site featuring a species-rich mix of native grasses and forbs, which were growing on relatively N-poor soils. A decade of experimental N inputs to Field C saw declines in biodiversity, measured by species richness and the exponential of the Shannon   \\

\noindent\fbox{%
	\parbox{\linewidth}{%
		\textbf{Box 1: Defining and Measuring Biodiversity}\\
	
		Most references cited in this introductory section use species richness as a sole or primary measure of biodiversity, though some consider additional dimensions of biodiversity such as functional group richness or evenness.
		
In this work, we use the abundance of native grasses as a proxy for biodiversity in a model of a nutrient-enriched and invaded grassland  (see equations (\ref{eq:ode-system}a-g)). In North American grasslands, this modeling choice is consistent with observed correlations between native dominance and species richness as well as Simpson diversity  \cite{martin2014biodiversity,wilsey2011biodiversity}. }
}\\ 

\noindent diversity index ($e^H$) \cite{clark2008loss,isbell2013low}. Meanwhile, the exotic European  grass \emph{Elymus repens} increased in abundance \cite{isbell2013low}. Experimental N inputs ceased in 1992 in a subset of the replicate plots, and soil N recovered to control levels \cite{isbell2013low}. However, more than two decades later, species richness and $e^H$ remained low and exotic abundance remained high in plots that had received the highest levels of N inputs \cite{isbell2013low}. The failure of the plant community to return to a pre-fertilization state after a cessation of N inputs may indicate hysteresis has occurred.

At the Park Grass experiment, species richness similarly declined during a period of fertilization and N deposition, which lasted over a century (\cite{silvertown2006park} and references therein). But when the experimental inputs ceased in 1989 and ambient rates of N deposition began to decrease, species richness, $e^H$, and the Simpson diversity index increased considerably within a decade \cite{storkey2015grassland}. The reversibility of biodiversity loss at Park Grass is consistent with the R* theory of resource competition \cite{tilman1982resource} in the context of a single limiting resource.

The fact that biodiversity has not recovered in this way at Cedar Creek invites further theoretical explanation.  \cite{storkey2015grassland} note a difference between the Park Grass and Cedar Creek experiments: sites at Park Grass were mowed twice yearly, removing plant biomass. In contrast, Cedar Creek sites were not mowed and invaded plots accumulated dense litter mats, which have been hypothesized to reinforce exotic dominance \cite{isbell2013low}. To explore the mechanisms by which plant litter may suppress the recovery of biodiversity, we develop a mathematical model of nitrogen, plant, and litter dynamics in which plants compete for N and produce litter that inhibits plant growth directly in a species-dependent manner. 
The construction of this model departs from classical plant-soil feedback models by explicitly representing resource (N) competition in addition to inhibitory (litter) interactions across species. It also
advances beyond a competitive Lotka-Volterra framework by resolving cross-species interactions mediated by N and litter as separate processes.
The presence of both resource competition and inhibitory mechanisms in the model support detailed analysis and predictions of the conditions for recovery after a resource pulse---both in grasslands and more generally in ecosystems that feature similar mechanisms.

The organization is as follows. In section ``Review of Mechanisms",
we discuss biological mechanisms that could prevent biodiversity from recovering at Cedar Creek and identify mathematical features that would represent non-recovery in a model of biodiversity dynamics. In section ``Model'', we introduce the compartmental model of competition between native and exotic species groups that incorporates competition for N, immigration, and litter inhibition of plant growth.  The section ``Model Behavior'' shows that the model can produce bistability and hysteresis in response to N loading under parameter values tabulated specifically for Cedar Creek. 
In section ``Conditions for bistability and hysteresis" we develop broader analytic criteria on parameters---such as those controlling inhibition and resource-dependent growth rates---for bistability and hysteresis, and we connect these criteria to empirical studies.  The ``Discussion'' section summarizes our findings, highlights the unique contributions of the model presented, and considers its implications both for biodiversity recovery in grasslands and for the occurrence of hysteresis in ecosystems more broadly.

\section*{Review of Mechanisms}
\label{sec:mech}

Potential mechanisms underlying the non-recovery of biodiversity following N cessation at Cedar Creek may be described biologically in terms of organic and inorganic agents. They can also be described mathematically in terms of the dynamic structures in models of species' interactions. We review biological and mathematical mechanisms in turn. 

\subsection*{Biological impediments to recovery}

Positive plant-soil feedback (PSF) \cite{bever1997incorporating,bennett2019mechanisms} favoring \textit{E. repens}  could in theory block recovery of native biodiversity at Cedar Creek. On the one hand, no evidence for microbial PSF has emerged from reciprocal inoculation experiments involving  mycorrhizal fungi mutualists from Field C at the Cedar Creek cessation experiment  \cite{portales2020}. 
On the other hand, a broader interpretation of PSF encompasses feedbacks between plants and abiotic soil conditions such as nutrient content and secondary chemicals; PSF can also be modulated by temperature and moisture variation \cite{bennett2019mechanisms,delong2019plant}. 

Plant litter has the potential to impact these abiotic factors, and empirical studies across several North American grassland sites suggest that plant litter controls biodiversity. In an Alberta fescue grassland, high litter mass predicted low species richness and evenness \cite{lamb2008direct}. Litter manipulations both at a Michigan midsuccessional old field site \cite{foster1998species} and at Cedar Creek \cite{clark2010recovery} detected significant negative effects of litter mass on species richness.  Furthermore, because fires and grazers such as bison historically prevented dead biomass from accumulating in North American tallgrass prairies, native species in these communities may have rarely experienced high litter conditions \cite{knapp1986detritus}. Results at the Cedar Creek N addition and cessation experiment align with these observations: litter accumulated markedly in the nutrient-enriched, invaded plots and correlated negatively with native species richness \cite{isbell2013low}.

One way in which plant litter might impede the recovery of native biodiversity is by altering nutrient cycling.
If accumulated N recycles through litter, soil, and plant tissue, then the legacy of experimental inputs could persist long after their cessation \cite{tilman2015biodiversity}. 
In contrast to sediments' role in simple models of lake eutrophication \cite{carpenter1999management}, litter decomposition is not expected to respond nonlinearly to increasing N availability.
Nonetheless, litter quality and decomposition rates can feed back positively on fast-growing species \cite{bennett2019mechanisms}, and might reinforce dominance of a few such as \textit{E. repens}. Given the centrality of N as a limiting and modified resource for grasslands in general and Cedar Creek in particular, we base equations (\ref{eq:ode-system}) (section ``Model") on N cycling through soil, plant, and litter pools.

Plant litter could also impede native biodiversity recovery through a number of additional mechanisms that we will refer to collectively as litter inhibition. Litter might release allelopathic chemicals such as those detected in \textit{E. repens} shoots  \cite{weston1987isolation}.
In addition, a litter layer can intercept soil-bound seeds and hinder seedlings from emerging mechanically \cite{facelli1991plant,foster1998species}. Litter also lowers light availability at the soil surface, which can suppress seed germination and establishment.
 Existing models of exotic plant invasions suggest that light interception by invaders \cite{chisholm2015potential} or particuarly their litter \cite{eppinga2011litter} prevents recovery of natives following a nutrient pulse. 
We represent a broader class of possible litter inhibition mechanisms in  equations (\ref{eq:ode-system}) (section ``Model") by specifying plant growth rates as decreasing functions of litter stocks.

In addition to PSF and litter effects, grassland seed limitation \cite{foster2003seed,foster2007restoration} is another possible obstacle to native community recovery. In subplots of Field C at Cedar Creek, seed additions in fall and spring of 2004 significantly increased species richness the following year \cite{clark2010recovery}. On the other hand, Field C receives propagules from an adjacent high-diversity field of native perennials. If seed limitation were at play, we would expect a correlation between the proximity of experimental plots to this propagule source and the plot's species richness; however, none was found \cite{isbell2013low}. 
To further explore the effect of seed and establishment limitation, we include terms that represent plant immigration via external seed supply in equations (\ref{eq:ode-system}) (section ``Model") and vary the immigration rate in our analysis.

\subsection*{Mathematical structures of non-recovery}

The persistence of low diversity at Cedar Creek could in theory reflect either an alternative stable state or a long transient state \cite{hastings2018transient} as diversity recovers imperceptibly slowly on the decadal timescales of the experiment \cite{francis2021management,isbell2013low}. 
Here we focus on mechanisms sufficient to produce two alternative stable states at identical N input rates, namely a species-rich community of mostly native plants and a species-poor community dominated by exotics.

\begin{figure}[h!]
    \centering
    \includegraphics[width=0.5\linewidth]{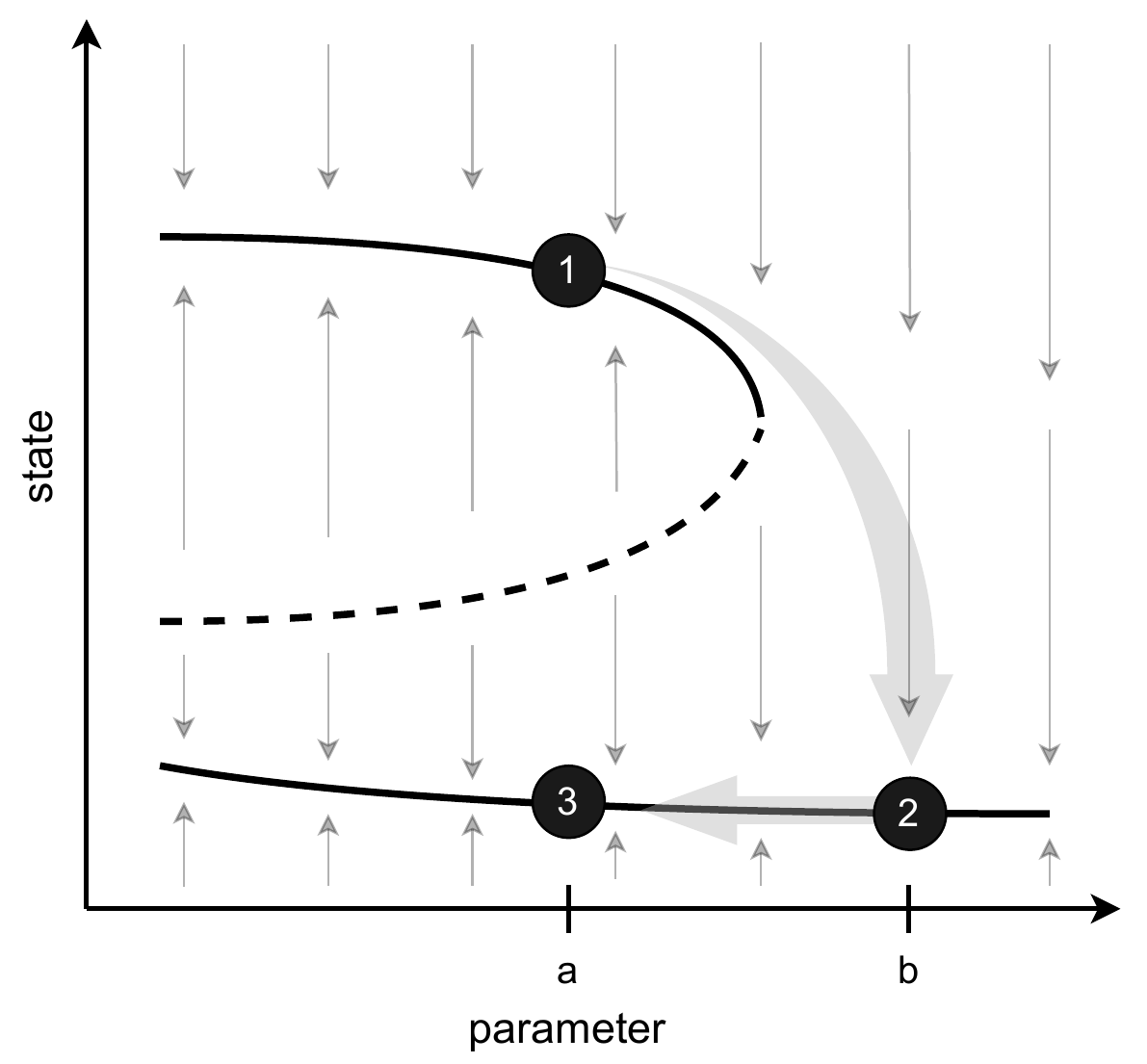}
    \caption{
    \
    \small {\bf An example of bistability and hysteresis.} Vertical arrows indicate dynamic changes in state for a fixed parameter. Solid and dashed lines are stable and unstable branches of equilibria, respectively. At parameter value $a$, the system is bistable because it can stabilize at point 1 or 3, depending on its initial state. If it starts at point 1 and the parameter increases from $a$ to $b$, the state shifts to point 2 on the lower branch of equilibria (long grey arrow). If the parameter then returns from $b$ to $a$, the state transitions to point 3 (short grey arrow). The state of the system depends on parameter history, so the system is hysteretic.}
    \label{fig:generic_hysteresis}
\end{figure}

The potential for alternative stable states to structure ecological systems has been studied since the late 1960's, with mathematical models playing an important role \cite{beisner2003alternative, schroder2005direct}. In the language of dynamics, alternative stable states occur when multiple attractors exist for the same parameter values. When there are two such attractors, the system is `bistable'. Multiple attractors can enable hysteresis---dependence of system state on parameter history---as illustrated in Figure \ref{fig:generic_hysteresis}. 

Suppose a parameter value begins at $a$ and the state equilibrates near point 1 on the upper stable branch of equilibria. If the parameter value then increases to $b$, the state will tend towards the lower branch of equilibria at point 2. Returning the parameter to $a$ does not restore the state to point 1; instead, it remains on the lower branch of equilibria at point 3. In this way the state of the system depends on not only the current parameter value but also its history---the hallmark of hysteresis. 

If one interprets the state variable in Figure \ref{fig:generic_hysteresis} as (native) species richness and the parameter as the N input rate, the behavior from point 1 to 2 to 3  qualitatively resembles the experimental findings at Cedar Creek. We turn to the question of whether species-dependent litter inhibition could generate such bistability and hysteresis in a model of grassland dynamics.   

\section*{Model}
\label{sec:model}
\begin{figure}[h!]
	\begin{center}
	\includegraphics[width = 0.75\linewidth]{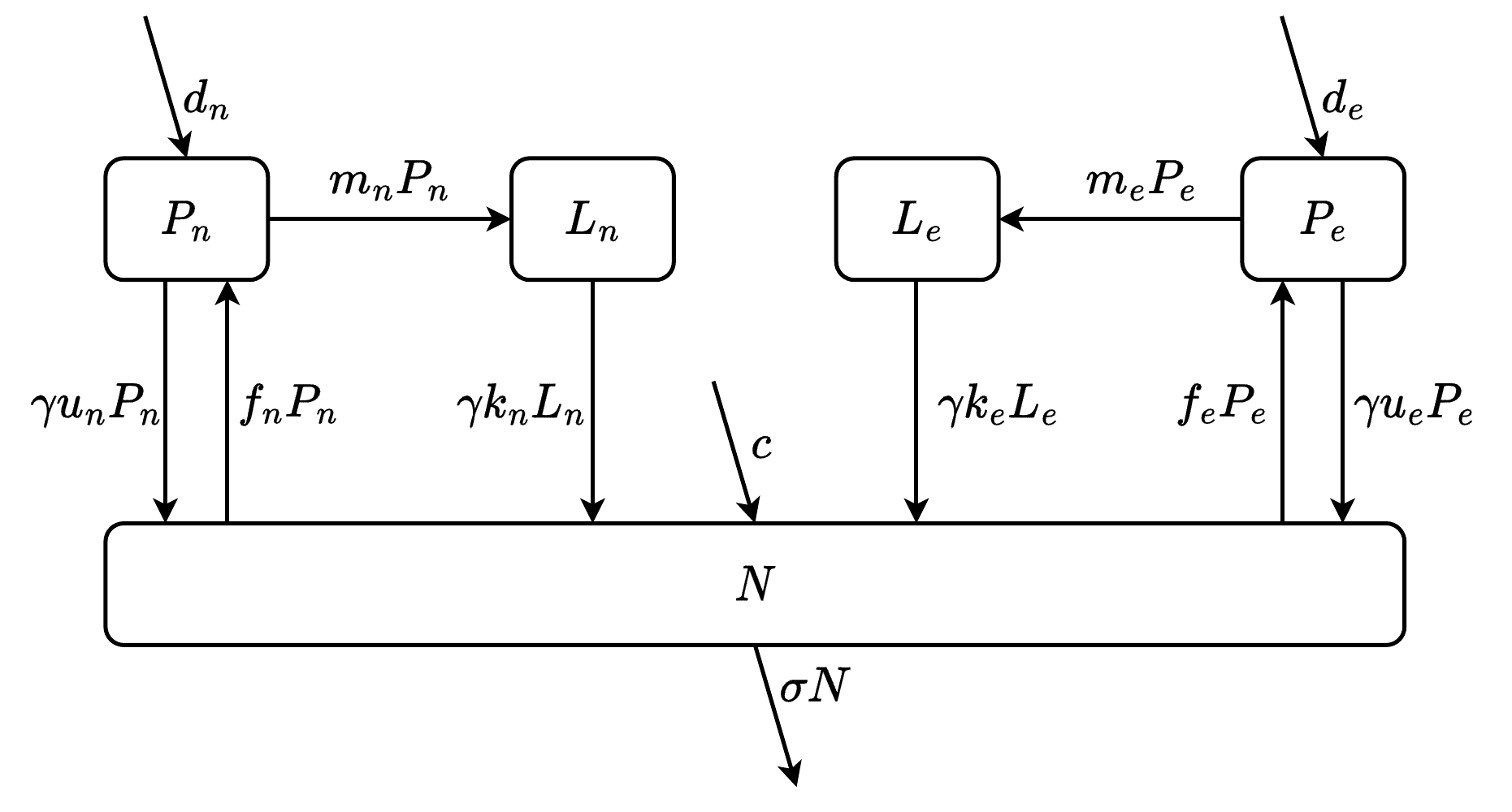}
	\end{center}
	\caption{\small {\bf Compartmental diagram representing the ODE system (\ref{eq:ode-system}).} Boxes represent state variables; arrows indicate changes in those state variables. See Table \ref{table:parameter-values} for details.}
	\label{compartmentalmodel}
\end{figure}

We explore the effects of nutrient cycling, immigration, and litter feedbacks on community dynamics within a compartmental model, illustrated in Figure \ref{compartmentalmodel}. The fluxes in model equations (\ref{eq:ode-system}) capture the cycling of nitrogen between soil ($N$), plant biomass ($P_i$), and litter mass ($L_i$).  Atmospheric deposition and experimental fertilization add to the soil inorganic N pool while leaching removes N. Nitrogen cycles from inorganic forms in the soil to organic forms in plant biomass and litter, and back to the soil inorganic pool via remineralization.  

We split each organic box (plant biomass $P_i$ and litter mass $L_i$) to track two groups of species. The native group tracked by $P_n$ and $L_n$ represents a diverse mix of grassland species present at the onset of N additions at Cedar Creek, typified by \textit{Schizachyrium scoparium}. The exotic group tracked by $P_e$ and $L_e$ represents a pair of cool-season European grasses---\textit{Elymus repens} and \textit{Poa pratensis}---that became dominant during N enrichment and persisted following its cessation. In many grasslands throughout the Great Plains, native dominance is correlated with higher species richness and Simpson diversity index \cite{martin2014biodiversity,wilsey2011biodiversity}.
Accordingly, in this paper ``native-dominated" and ``high-diversity" will be used interchangeably to describe plant communities, as will ``exotic-dominated" and "low-diversity."  This link between biodiversity and the relative abundance of native species aligns with a call by \cite{hillebrand2018biodiversity} for biodiversity indices that are sensitive to shifts in species identities. It also has the mathematical advantage of limiting the number of variables needed to represent the state of the system.

The resulting model has five state variables: native and exotic plant biomass $P_n$ and $P_e$, native and exotic litter mass $L_n$ and $L_e$, and inorganic soil nitrogen $N$:

\begin{subequations}\label{eq:ode-system}
\begin{linenomath*}
\begin{align}
    \frac{dP_n}{dt} &= \left(f_n(L_n,L_e,N) - m_n - u_n \right)P_n + s_n \label{eq:ode-system_a}\\
    \frac{dP_e}{dt} &=\left(f_e(L_n,L_e,N)- m_e - u_e \right)P_e + s_e \label{eq:ode-system_b}\\
    \frac{dL_n}{dt} &= m_n P_n - k_nL_n \label{eq:ode-system_c}\\
    \frac{dL_e}{dt} &= m_e P_e- k_eL_e \label{eq:ode-system_d}\\
    \frac{dN}{dt} &= c - \sigma N +\gamma\Big(k_n L_n + k_e L_e +u_n P_n + u_e P_e  -f_n(L_n,L_e, N) P_n - f_e(L_n,L_e, N)P_e\Big) \label{eq:ode-system_e}\\
    &\text{where} \hspace*{1cm} f_n(L_n,L_e,N)=\frac{g_n N}{h_n + N}e^{-\beta_{nn} L_n  -\beta_{en} L_e } \label{eq:ode-system_f}\\
    &\text{and} \hspace*{1.5cm} f_e(L_n,L_e,N)=\frac{g_e N}{h_e + N} e^{-\beta_{ne} L_n-\beta_{ee} L_e} \label{eq:ode-system_g}
\end{align}
\end{linenomath*}
\end{subequations}   

This model uses linear terms to represent the rates of litter production ($m_iP_i$), litter decomposition ($k_iL_i$), belowground death ($u_iP_i$), and  leaching of inorganic N from the soil out of the system  ($\sigma N$). For simplicity, belowground death returns N to the soil inorganic pool immediately. Constant terms represent inputs from outside the system. The parameter $c$ adds inorganic N from atmospheric deposition and/or experimental manipulation. 
The parameters $s_n$ and $s_e$ represent plant immigration due to dispersal and establishment of native and exotic seeds, respectively. 
The functions $f_n$ and $f_e$ model nonlinear effects of soil inorganic N and litters $L_i$ on per-capita plant growth rates. The first factor in each, $r_i(N)=\dfrac{g_i N}{h_i+N}$, represents a type-II response of plant growth to N availability, with half saturation constant $h_i$ and asymptotic value $g_i$. The remaining factors depress this growth rate based on native and exotic litter abundances; parameters $\beta_{ij}$ control the sensitivity of plant group $j$ to litter from group $i$\footnote{While the structure of intra- and inter-specific feedback parameters $\beta_{ij}$ is reminiscent of classical microbial PSF models (e.g. \cite{bever1997incorporating}), the inclusion of nutrient cycling in the model (\ref{eq:ode-system}) opens the possibility to analyze outcomes under explicitly changing nutrient availability.}. The constant $\gamma$ in equation (\ref{eq:ode-system_e}) converts plant tissue mass to mass of inorganic soil nitrogen. Tissue nitrogen concentrations vary according to many factors including species and decomposition stage \cite{risser1982ecosystem,pastor1987little}; our use of a single factor $\gamma$ represents a modeling simplification.
Table \ref{table:parameter-values} (at the end of this manuscript) summarizes the parameters of system (\ref{eq:ode-system}). 

\linespread{1.3}
\begin{sidewaystable}[h!]
\begin{small}
\caption{State Variables and Parameters of Model (\ref{eq:ode-system})}
\label{table:parameter-values}
\centering
	\begin{tabular}{@{}cllll@{}}
		\toprule
		State Variable & Description & Units \\
		\midrule
		$P_n$   & native plant biomass & g m$^{-2}$ \\
	    $P_e$   & exotic plant biomass & g m$^{-2}$ \\
	    $L_n$   & native plant litter & g m$^{-2}$ \\
	    $L_e$   & exotic plant litter & g m$^{-2}$ \\
	    $N$     & soil inorganic N & g m$^{-2}$ \\
	    \midrule
		Parameter & Controls... & Units & Default Value(s)\\
	    \midrule
		$c$ & rate of atmospheric and experimental N deposition & g m$^{-2}$ yr$^{-1}$ & 1-10  \\
	    $\sigma$ & rate of inorganic N leaching & yr$^{-1}$ & 1  \\
	    $\gamma$ & N content of plant tissue & g g$^{-1}$ & 0.01 \\
		$s_n$, $s_e$ & rates of native, exotic immigration & g m$^{-2}$ yr$^{-1}$ & 0.1  \\
		$m_n$, $m_e$ & rates of native, exotic litter production & yr$^{-1}$ & 0.2 \\
		$u_n$, $u_e$ & rates of native, exotic underground plant tissue senescence  & yr$^{-1}$ & 0.2  \\
		$k_n$, $k_e$ & rates of native, exotic litter decomposition & yr$^{-1}$ & 0.2  \\
		$g_n$ & native per-capita growth rate when $N$, litter not limiting & yr$^{-1}$ & 200  \\
		$g_e$ & exotic per-capita growth rate when $N$, litter not limiting & yr$^{-1}$ & 250  \\
		$h_n$ & half-saturation point in native growth rate response to $N$  & g & 0.2  \\
		$h_e$ & half-saturation point in exotic growth rate response to $N$ & g & 1.5   \\
		$\beta_{nn}$ & native litter inhibition of native plant growth & g$^{-1}$ & 0.05  \\
		$\beta_{en}$ & exotic litter inhibition of native plant growth & g$^{-1}$ & 0.1  \\
		$\beta_{ne}$ & native litter inhibition of exotic plant growth & g$^{-1}$ & 0.05 \\
		$\beta_{ee}$ & exotic litter inhibition of exotic plant growth & g$^{-1}$ & 0.05 \\
		\bottomrule
	\end{tabular}
\bigskip{}
\\
{\footnotesize Values of $c$ were chosen in reference to the National Atmospheric Deposition Program dataset (\url{http://nadp.slh.wisc.edu/data/sites/siteDetails.aspx?net=NTN&id=MN01}) and fertilization treatments reported by \cite{isbell2013low}. Values of $\sigma$, $m_i$, and $u_i$ were chosen to achieve order-of-magnitude agreement with Cedar Creek BioCON data, available via Experiment e141 at \url{https://www.cedarcreek.umn.edu/research/data}. A value for $\gamma$ was estimated based on results of \cite{risser1982ecosystem,pastor1987little}.} $k$ was estimated based on results of \cite{brandt2010role, pastor1987little, zhang2008rates}.  
\end{small}
\end{sidewaystable}
\linespread{1.7}

\section*{Model Behavior} 
\label{sec:behavior}
We illustrate bistability and hysteresis in the model (\ref{eq:ode-system}) using parameter values tailored to the Cedar Creek grassland community (see Table \ref{table:parameter-values}). For further exploration of parameter space and connection to empirical studies outside Cedar Creek, see section ``Conditions for bistability and hysteresis."

\subsection*{Bistability}

The model (\ref{eq:ode-system}) can produce bistability at intermediate N input levels, an important prerequisite for hysteresis in response to N loading. Bistability occurs in a parameter regime in which native plants have an N-dependent growth advantage at low soil N that reverses with fertilization ($h_n=0.2$, $h_e=1.5$, $g_n=200$, $g_e=250$) and exotic litter inhibits native species growth disproportionately ($\beta_{en}>\beta_{ne}$). Both conditions are biologically feasible at Cedar Creek,
as the dominant native grasses are known to be superior exploitative N competitors (e.g., \cite{tilman1991plant}) and to be limited in productivity by the accumulation of detritus \cite{knapp1986detritus}. 

\begin{figure}[h!]\label{fig:dispN}
    \centering
    \includegraphics[scale=0.58]{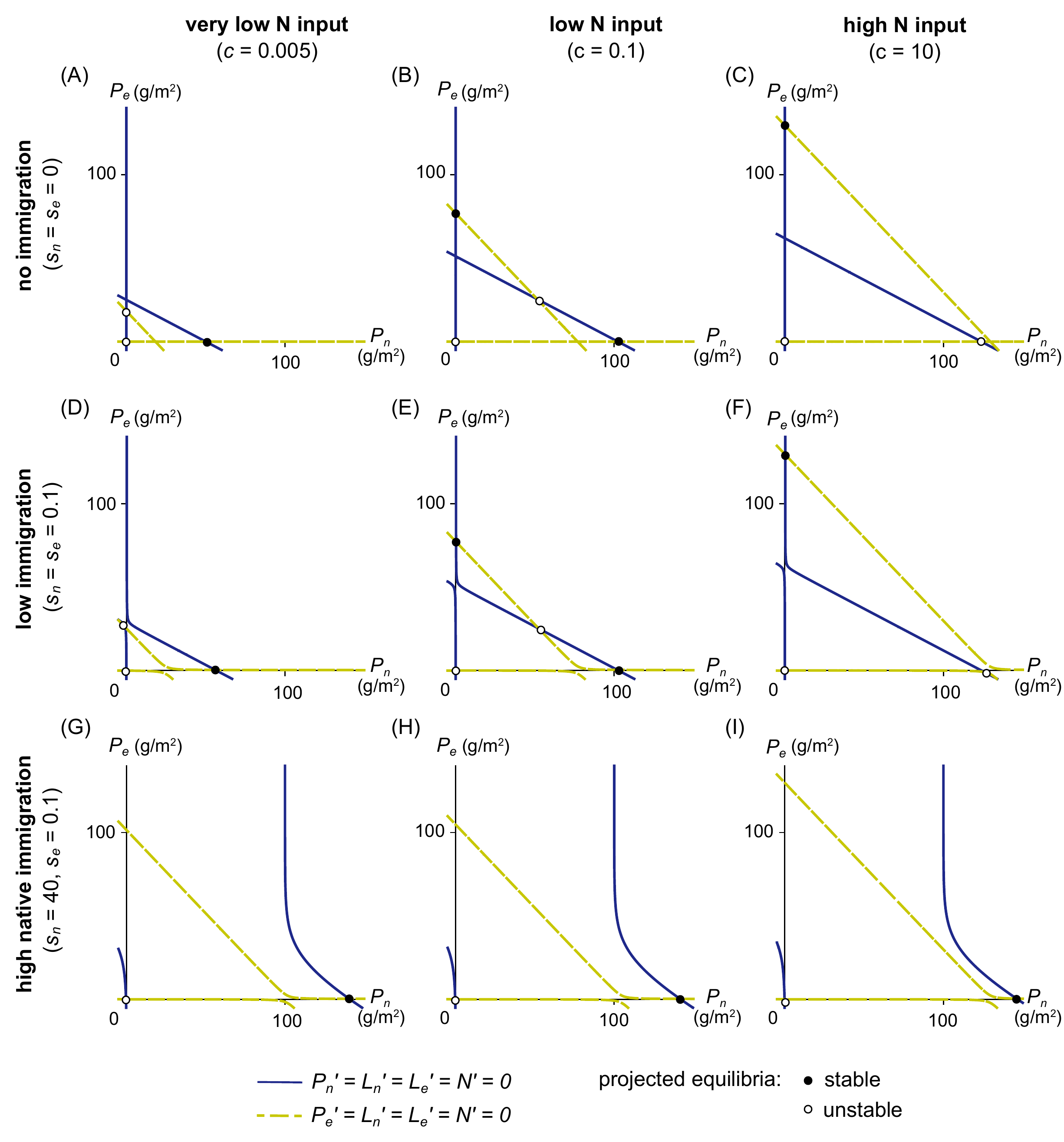}
    \caption{\small {\bf Bistability of ODE system (\ref{eq:ode-system}) for varying N inputs  and immigration rates}. Each of the blue (green) curves shown in the $P_n$,$P_e$ plane represents no-change conditions on all variables except $P_n$ ($P_e$, respectively). The intersection of the curves gives $P_n$,$P_e$ coordinates of the equilibria.  
    For zero or small immigration, the stable equilibria (filled circles) change from a single, native-dominated equilibrium at very low N inputs (panels A, D) to a pair of native- and exotic-dominated equilibria at low N inputs (panels B, E) and to a single exotic-dominated equilibrium at high N inputs (panels C, F). High immigration rates can destroy this structure; in particular, large values of the native immigration parameter $s_n$ can overcome exotic dominance (Panels G-I). Parameters are as in Table \ref{table:parameter-values}, except $c,s_n,s_e$ as indicated and $g_e=300$ to more clearly illustrate intersections between curves.}
    \label{fig:nullclines}
\end{figure}

To visualize bistability in the five-dimensional system, Figure \ref{fig:nullclines} projects to the $P_n$,$P_e$-plane. Equilibrium values of $N$, $L_n$, and $L_e$ can be recovered from values of $P_n$ and $P_e$ because at any  equilibrium point $(P_n^*,P_e^*,L_n^*,L_e^*,N^*)$, soil inorganic N is determined by the balance between

\noindent inputs and leaching ($N^*=c/\sigma$) and litter biomass is proportional to its respective plant biomass ($L_i^*=m_iP_i^*/k_i$) (see Appendix). Equilibria therefore occur at the  intersection between two types of curves: one on which all state variables but $P_e$ must have zero rate of change (dark blue in Figure \ref{fig:nullclines}), and another on which all state variables but $P_n$ must have zero rate of change (light green in Figure \ref{fig:nullclines}).  Under the simplifying assumption of negligible immigration ($s_i=0$), these curves are straight lines reminiscent of Lotka-Volterra nullclines (Figure \ref{fig:nullclines}A-C). Linear stability analysis of the full five-dimensional system confirms what the nullcline analogy suggests:  only the native-dominated equilibrium is stable at low N inputs (panel A), bistability occurs between the native- and exotic-dominated equilibria at intermediate N inputs (panel B), and only the exotic-dominated equilibrium is stable at high N inputs (panel C). The dynamic structures in panels A-C persist for small immigration rates (Figure \ref{fig:nullclines}D-F), but can be destroyed by sufficiently large immigration rates. For example, high native immigration results in a single stable equilibrium dominated by native plants across low, intermediate, and high N input levels (Figure \ref{fig:nullclines}G-I). Based on empirical evidence of seed and dispersal limitation in grassland communities \cite{foster2007restoration,foster2003seed,sullivan2018mechanistically}, we focus our attention in the present work on a low-immigration regime.

\subsection*{Hysteresis}

\begin{figure}[h!]
    \centering
    \includegraphics[width=0.9\linewidth]{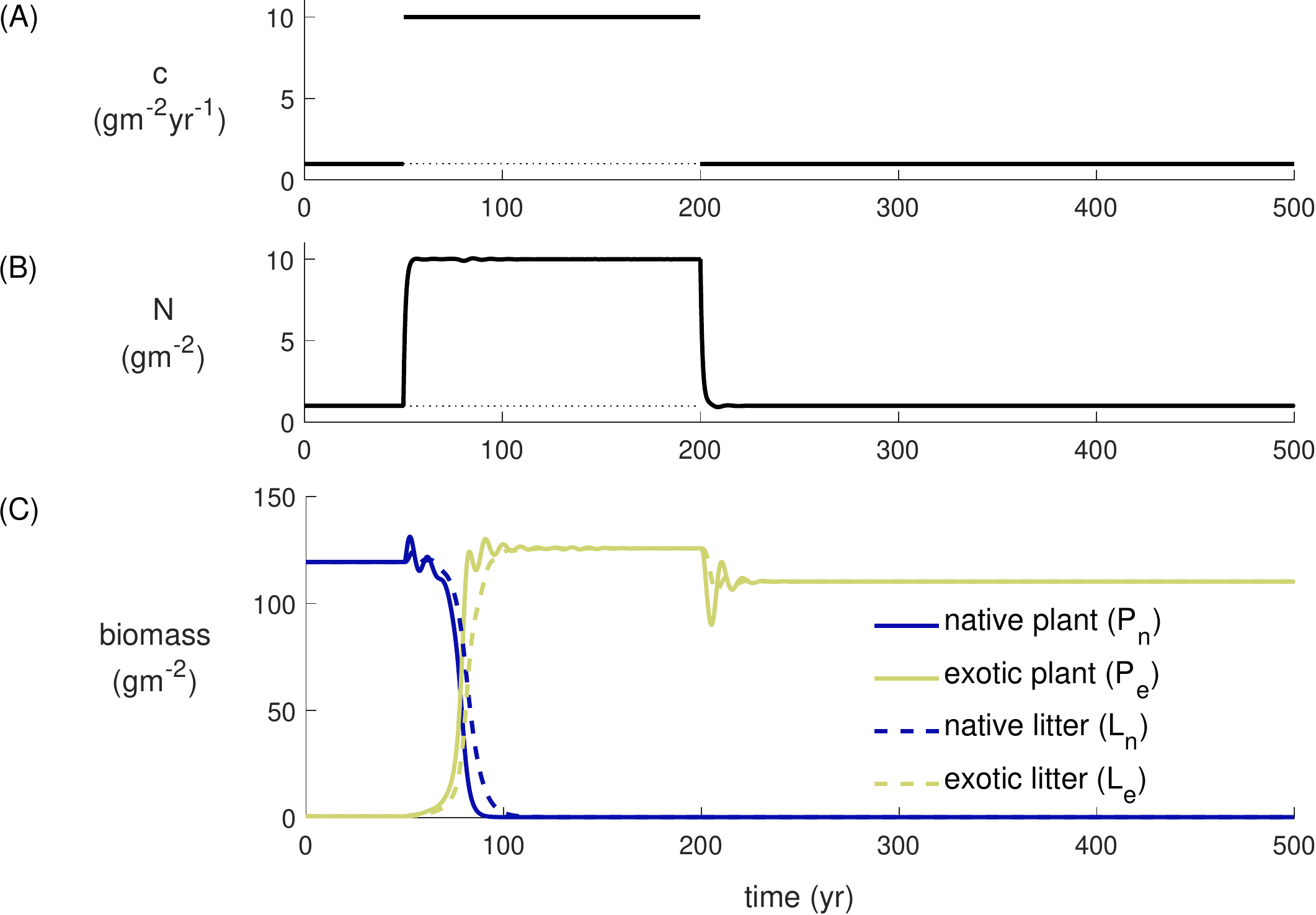}
    \caption{\small {\bf Hysteretic response of native and exotic populations to N addition and cessation.} (A) N inputs $c$ increase from 1 to 10 g m$^{-2}$ yr$^{-1}$ at 50 years and return to 1 g m$^{-2}$ yr$^{-1}$ 150 years later. (B) Soil inorganic N ($N$) re-equilibrates rapidly in response to changing input rates. (C) Native plants initially dominate in historic $N$ conditions but lose their dominance to exotic plants and litter under elevated $N$ during time 50-200 and fail to recover even when $N$ returns to original levels after time 200. 
    Parameters are as in Table \ref{table:parameter-values}. Timeseries for the ODE (\ref{eq:ode-system}) were generated using ode45 in MATLAB R2021a.}
    \label{fig:model-hysteresis}
\end{figure}

In the low-immigration regime of the model (\ref{eq:ode-system}), bistability can drive a hysteretic response of plant community composition to N addition and cessation that qualitatively mirrors the experimental results at Cedar Creek reported by \cite{isbell2013low}. Figure \ref{fig:model-hysteresis} illustrates this hysteretic behavior in a timeseries simulated by the solver ode45 in MATLAB R2021a. During the lower (higher) N-input phases of the timeseries, nullclines qualitatively resemble Figure \ref{fig:nullclines}E (F, respectively). As in the Cedar Creek experiment, N inputs (Figure \ref{fig:model-hysteresis}A) and soil N content (Figure \ref{fig:model-hysteresis}B) begin at historic levels. The biodiverse group of native plants (solid, dark blue line) initially dominates over the exotic, low-diversity group (solid, light green line) (Figure \ref{fig:model-hysteresis}C). A ten-fold increase in N inputs at $t=50$ years causes soil N to rapidly re-equilibrate at an elevated level (panel B) and exotic species to replace natives (panel C). When N inputs are reduced at $t=200$, soil N quickly returns to its original level, but the exotic species retain their dominance as they did at Cedar Creek.

The hysteretic behavior modeled in Figure \ref{fig:model-hysteresis} arises from differences between the native and exotic groups' growth responses to both N and litter. Differences in litter inhibition strengths ($\beta_{en}>\beta_{nn}$) contribute to bistability at intermediate N inputs (see subsection ``Conditions for bistability and hysteresis," below). Different growth responses to N including both asymptotic values ($g_e>g_n$) and half-saturation constants ($h_n<h_e$) reverse the native growth advantage as N inputs increase. Consequently, increasing N inputs shifts nullclines in favor of the exotic group as depicted in Figure \ref{fig:nullclines}E,F, prompting a transition to an exotic-dominated equilibrium. A biological interpretation of the modeled timeseries---also a plausible explanation for the Cedar Creek results---is as follows.
The native group initially maintains dominance because its N-efficiency advantage at historic soil inorganic N levels outweighs its litter disadvantage for low levels of exotic litter. When inorganic N inputs increase, the natives lose their low-N growth advantage. 
Although the exotics are at low abundance initially, their N-dependent growth advantage and the disproportionate effect of their litter on native growth allows them to displace native species over time.
Upon return of inorganic N inputs and soil levels to original levels, the high quantity of exotic litter counteracts the native growth advantage at low N and exotics maintain abundance.

\section*{Conditions for Bistability and Hysteresis}
\label{sec:conditions}
In this section we derive mathematical conditions on the parameters in the model (\ref{eq:ode-system}) that are necessary to produce bistability and hysteresis. We interpret these conditions biologically and discuss settings in which the model may predict dynamics.

\subsection*{Conditions for Bistability}

Bistability in the model (\ref{eq:ode-system}) occurs under biologically meaningful conditions on parameters controlling plant growth and litter dynamics.
To allow analytic derivation of these conditions we employ the simplifying assumption that no immigration occurs. This simplification is reasonable in the sense that bistability that exists for $s_i=0$ persists for small $s_i$ (see Figure \ref{fig:nullclines}B,E); however, it should be noted that not all bistable scenarios for $s_i>0$ may be captured by the analysis that follows.

First, any equilibrium of interest must feature non-negative population densities. Native- and exotic-dominated equilibria exist at positive population densities exactly when 
\begin{subequations}\label{eq:pos-cond}
\begin{linenomath*}
\begin{align}
    r_n(N^*)&>m_n+u_n \label{eq:pos-conda} \\
    \text{and}\hspace{1cm} r_e(N^*)&>m_e+u_e, \label{eq:pos-condb}
\end{align}
\end{linenomath*}
\end{subequations}
\noindent where $r_i(N) =\frac{g_i N}{h_i+N}$ denotes N-dependent plant growth rates and $N^* =\frac{c}{\sigma}$ denotes equilibrium soil N. These conditions require each group's per-capita growth rate under equilibrium soil N and litter-free conditions $(r_i(N^*))$ to exceed its  per-capita mortality $(m_i+u_i)$. Litter decomposition rates and inhibition strengths do not alter the existence of biologically relevant native- and exotic- dominated equilibria.

In contrast, litter processes do impact the location and stability of the native- and exotic-dominated equilibria.  Linear analysis (described in the Appendix) yields the following conditions necessary for their mutual stability: 
\begin{equation}\label{eq:stab-cond}
\begin{linenomath*}
    \frac{\beta_{nn}}{\beta_{ne}} <\frac{\ln\left(\frac{r_n(N^*)}{m_n+u_n}\right)}{\ln\left(\frac{r_e(N^*)}{m_e+u_e}\right)}  
    <\frac{\beta_{en}}{\beta_{ee}}
\end{linenomath*}
\end{equation}
\noindent where, as before, $r_i(N)=\frac{g_i N}{h_i+N}$ and $N^*=\frac{c}{\sigma}$. The left (respectively, right) inequality in (\ref{eq:stab-cond}) is necessary for the stability of the equilibrium dominated by native (respectively, exotic) plants. The sufficiency of these conditions for stability is unknown (see Appendix).

Each ratio in (\ref{eq:stab-cond}) can be interpreted as an advantage or disadvantage of native species relative to exotics, while the inequalities describe balances among these advantages and disadvantages.
For example, the ratios $\beta_{in}/\beta_{ie}$ measure the inhibitory effect of litter from species group $i$ on native growth relative to exotic growth; a ratio greater than one represents an inhibitory disadvantage for the natives.  
In the middle term, the input to each logarithm $\frac{r_i(N^*)}{m_n+u_n}$ is the ratio of per-capita growth and mortality rates for species group $i$ under equilibrium soil N availability and litter-free conditions.  
The full inequalities (\ref{eq:stab-cond}) can thus be interpreted as saying that bistability requires natives' relative litter-free growth strength $\left(\ln\left(\frac{r_n(N^*)}{m_n+u_n}\right)/\ln\left(\frac{r_e(N^*)}{m_e+u_e}\right)  \right)$ to exceed their relative inhibition by their own litter ($\beta_{nn}/\beta_{ne}$), but to be less than their relative inhibition by exotic litter ($\beta_{en}/\beta_{ee}$). The inequality (\ref{eq:stab-cond}) can also be interpreted as a bistability condition for superior and inferior resource competitors that produce inhibitory materials; we explore this viewpoint in the discussion.

Provided that native litter inhibits exotic growth no more than its own ($\beta_{ne}\leq\beta_{nn}$), the ratio $\beta_{nn}/\beta_{ne}$ in (\ref{eq:stab-cond}) is greater than or equal to one. In this scenario, stability of the native-dominated equilibrium requires via the first inequality in (\ref{eq:stab-cond}) that 
\begin{equation}
    \frac{r_n(N^*)}{m_n+u_n}>\frac{r_e(N^*)}{m_e+u_e}.  
\end{equation}

\noindent Our tabulated parameter values (see Table \ref{table:parameter-values}) from Cedar Creek's BioCON experiment predict order-of-magnitude agreement between the native and exotic species' mortality rates. The reduced inequality 
\begin{equation}\label{eq:simple-r}
    r_n(N^*)>r_e(N^*)
\end{equation}
suggests that in order to stabilize the native-dominated equilibrium, the native plants' intrinsic growth rate under prevailing inorganic nitrogen availability should exceed that of the exotics.

In alignment with inequality (\ref{eq:simple-r}), the native grass \emph{Schizachyrium scoparium} outcompeted both exotics \emph{Elymus repens} and \emph{Poa pratensis} within 5 years of co-seeding in unfertilized plots during a pairwise competition experiment at Cedar Creek \cite{wedin1993competition}. The low-N advantage of native bunchgrasses at Cedar Creek is consistent with the R* theory of resource competition \cite{tilman1982resource}, as monocultures of \emph{S. scoparium} as well as \emph{Andropogon gerardii} draw down soil inorganic N to significantly lower levels than monocultures of exotic competitors \cite{tilman1991plant,wedin1993competition}. 
This pattern of native advantage in low-N habitats
occurs more broadly in the central grasslands of North America \cite{vinton2006plant}, and also in western sagebrush-steppe communities \cite{vasquez2008creating}.

Rearranging inequalities (\ref{eq:stab-cond}) gives another condition necessary for bistability that is reminiscent of a two-species competitive Lotka-Volterra model:
\begin{equation}\label{eq:betas}
\begin{linenomath*}
    \beta_{nn}\beta_{ee}<\beta_{en}\beta_{ne}.
\end{linenomath*}
\end{equation}
In contrast to a Lotka-Volterra model, the cross- and self- inhibition processes parameterized by the $\beta_{ij}$ in model (\ref{eq:ode-system}) are mediated specifically through each group's litter. One way for inequality (\ref{eq:betas}) to hold is for native plants to be inhibited more by exotic litter than by native litter ($\beta_{en}>\beta_{nn}$) while exotic plants are not inhibited more by their own litter than by native litter ($\beta_{ee}\leq\beta_{ne}$).

Disproportionate inhibition of native plants by exotic litter may hold in North American tallgrass prairies in general and Cedar Creek's N-cessation experiment in particular. Native tallgrass prairie plants are certainly limited by the accumulation of litter, which fires and grazing historically removed---for example, litter hinders early-season growth of the native tallgrass \emph{A. gerardii} by blocking light and adding heat stress \cite{knapp1986detritus}.  In contrast, the cool-season European grasses \emph{P. pratensis} and \emph{E. repens} maintain high combined productivity at Cedar Creek despite litter accumulation: \emph{P. pratensis} is able to grow through the dense litter mats of \emph{E. repens} and the two cycle in dominance in invaded plots (\cite{isbell2013low} and subsequent observations).

While empirical evidence suggests that litter in general inhibits North American tallgrasses disproportionately relative to exotic invaders, inequality (\ref{eq:betas}) requires more.  The difference is subtle, but important: litter effects must be source-dependent. If, on the contrary, native and exotic litters had indistinguishable effects on plant growth, then the parameters $\beta_{ij}$---representing litter from species $i$ inhibiting the growth of species $j$---would satisfy $\beta_{nn}=\beta_{en}$ and $\beta_{ne}=\beta_{ee}$. This would imply that $\beta_{nn}\beta_{ee}=\beta_{en}\beta_{ne}$. To satisfy inequality (\ref{eq:betas}), the effect of litter on at least one species group must depend on which group shed the litter. As noted above, it suffices for native plants at Cedar Creek to be inhibited more by exotic litter than by native litter ($\beta_{en}>\beta_{nn}$). This hypothesized effect could stem from allelopathic compounds in \emph{E. repens} \cite{weston1987isolation} or from features of exotic litter that differentially intercept light, trap heat, or harbor pathogens. 

Source-dependent litter feedbacks on plant growth have been documented in a variety of ecosystems. For example, source-dependent effects have been found among annual and perennial grasses in a sage-brush steppe community in the Northern Great Basin, USA 
\cite{bansal2014plant}, between C$_4$ annuals native and invasive to northeast China \cite{li2016species}, among tropical trees in the Amazonian forest of French Guiana \cite{coq2012litter}, and between grass litter and forbs \cite{bosy1995mechanisms, chen2018mechanisms}.
In some systems, the presence of litter may facilitate rather than inhibit plant growth \cite{chapman2011away, hovstad2009conspecific, sarker2020species}. The effects of litter can also depend on environmental factors, such as temperature and precipitation, and may vary between greenhouses or shade houses and  field experiments \cite{olson2002effects, ruprecht2010differential, vincent2013major}. The nuances and variability in litter feedback effects across ecosystem types highlight the importance of site-specific measurements at Cedar Creek and elsewhere.
Targeted litter-exchange experiments could provide stronger evidence for or against bistability condition (\ref{eq:betas}) at Cedar Creek and other sites of interest.

\subsection*{Conditions for Hysteresis}
Bistability is necessary but not alone sufficient in our model to produce a hysteretic shift in plant community composition in response to N loading. Hysteresis also requires that bistability disappears at high levels of N loading, as in the transition in Figure \ref{fig:nullclines} from panel B to panel C. 

Mathematical explanations for N-induced destabilization of the native-dominated equilibrium in our model are aligned with empirical observations.  
Recall from (\ref{eq:stab-cond}) that the stability of native-dominated equilibrium depends on the relative magnitudes of $\beta_{nn}/\beta_{ne}$ and \\ $\ln\left(\frac{r_n(N^*)}{m_n+u_n}\right)/\ln\left(\frac{r_e(N^*)}{m_e+u_e}\right)$.
Under the simplifying assumptions that native litter inhibits each species group equally ($\beta_{nn}/\beta_{ne}=1$) and that natives and exotics experience the same per-capita mortality rates, the stability criterion reduces to comparing the intrinsic growth rates 
\begin{equation}\label{eq:stab-compare}
    r_n=\dfrac{g_n \frac{c}{\sigma}}{h_n+\frac{c}{\sigma}} \ \ \ \text{and} 
    \ \ \ r_e=\dfrac{g_e \frac{c}{\sigma}}{h_e+\frac{c}{\sigma}}.
\end{equation}
Here we have expanded the equilibrium nitrogen level as $N^*=c/\sigma$ to highlight the modeled dependence of the intrinsic growth rates on the nitrogen input rate $c$. The ordering of native and exotic intrinsic growth rates reverses in response to N loading, as shown in Figure \ref{fig:rnre}. When N inputs are low, $r_n>r_e$ promotes stability of the native-dominated equilibrium. When N inputs are high, $r_e>r_n$ destabilizes the native-dominated equilibrium. Parameter choices $g_e>g_n$ and $h_e> h_n$ encode this qualitative feature in the type-II responses in our model. 

Empirical evidence also suggests that increasing N inputs reduces native growth advantages at Cedar Creek and beyond. In N-fertilized plots at Cedar Creek, the exotic species  \emph{Elymus repens} and \emph{Poa pratensis} were able to reduce native \emph{Schizachyrium}'s biomass \cite{wedin1993competition}. More broadly, a worldwide NutNet study of 34 grassland sites spanning six continents found that nutrient enrichment increased exotic cover and decreased native richness \cite{seabloom2015plant}. Nutrient enrichment has also been implicated in invasions of New England tidal marshes by European \emph{Phragmites} 
\cite{holdredge2010nutrient}. These observations align with the prediction of the model (\ref{eq:ode-system}) that eutrophication can trigger a shift away from a native-dominated state. The model (\ref{eq:ode-system}) predicts a hysteretic response to cessation of N loading because the system can stay at an exotic-dominated stable equilibrium when N inputs reduce, as from panel C to panel B in Figure \ref{fig:nullclines}.\\

\begin{figure}[h!]\label{fig:rnre}
    \centering
    \includegraphics[width=0.9\linewidth]{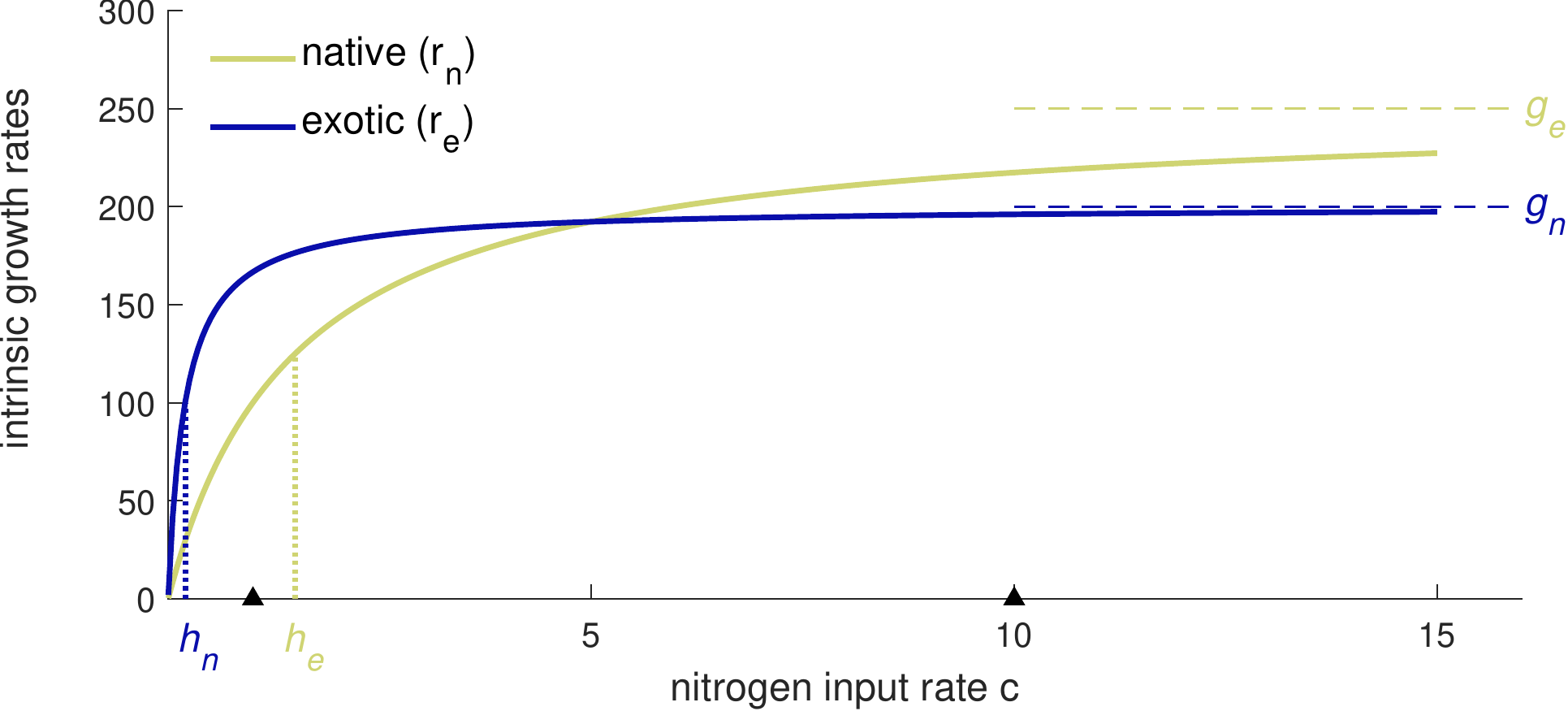}
    \caption{\small {\bf Intrinsic growth rates of native and exotic groups as N loading varies.} Intrinsic growth rates $r_i$ respond in a Type-II manner to N inputs (see equations (\ref{eq:stab-compare})). When exotics have a higher N-unlimited growth rate ($g_e>g_n$) and natives have a lower half-saturation constant ($h_n<h_e$), natives enjoy an intrinsic growth advantage at low N input rate $c$ that reverses at high N inputs. Triangles on the horizontal axis show $c=1$ and $c=10$ (compare to Figure \ref{fig:model-hysteresis}). Parameters are as in Table \ref{table:parameter-values}.}
    \label{fig:rnre}
\end{figure}

\section*{Discussion}
\label{sec:discussion}
To explore the possibility that accumulated litter has suppressed the recovery of native biodiversity following N-cessation at Cedar Creek, we have developed a model with the following features: (a) nitrogen cycles through organic and inorganic stores, added by atmospheric deposition or fertilization and removed by leaching; (b) native and exotic species groups grow in competition for available soil nitrogen; (c) native and exotic plants shed litter, which may accumulate before decomposing; (d) accumulated litter inhibits plant growth in a  species-dependent manner; and (e) immigration may occur from external propagule sources. Our mathematical analysis suggests that under low immigration rates, species-dependent litter inhibition can indeed stabilize multiple grassland community compositions, with their concomitant biodiversity levels. In particular, two stable states dominated alternately by natives (high diversity) or exotics (low diversity) occur when  native plants outcompete exotics under low-N, low-litter conditions but exotic litter disproportionately inhibits native plant growth (see inequalities (\ref{eq:stab-cond})). 
These conditions for bistability in community composition are plausible both at Cedar Creek and more broadly in North American grasslands, suggesting that reversing N-induced shifts to these systems may require active management.

Burning, grazing, and other methods of biomass removal are commonly employed in tallgrass prairie restoration \cite{wagle2018tallgrass}.
During experimental interventions conducted at Cedar Creek in 2004 and 2005 \cite{clark2010recovery}, raking away litter boosted seedling establishment and plant diversity, particularly when coupled with seed additions. 
At Park Grass, annual haying may have helped biodiversity to recover by countering the accumulation of litter. 

To complement field studies, the presented model offers a unified theoretical framework for exploring these potential restoration strategies---for example, by changing parameters controlling N addition ($c$) and native immigration ($s_n$) or by manipulating the state variables of biomass ($P_i$) and litter ($L_i$). 
Based on Figure \ref{fig:nullclines}A,D, extreme reductions in $c$ could in theory destabilize the exotic-dominated state and allow native plants to recover. While these reductions may be difficult to achieve in practice through curtailment of anthropogenic sources alone, organic carbon additions \cite{blumenthal2003soil} and controlled burns may help remove available N, effectively lowering $c$.  
Alternatively, high levels of native immigration $d_n$ in Figure \ref{fig:nullclines}G-I eliminates the exotic-dominated equilibrium entirely, suggesting that seed additions may also help overcome exotic dominance. 
Even when a stable exotic-dominated equilibrium persists, direct manipulation of plant biomass and/or litter (e.g. via grazing, haying, burning) may promote recovery towards the native-dominated equilibrium by shifting the system state into the latter's basin of attraction. 
Effective restoration may require a combination of the above  strategies, and the model may be used to explore potential combinations, informing the design of empirical studies. 
The model's utility for predicting specific restoration strategies stems from explicitly representing multiple targets for management: in particular, the nitrogen for which plants compete and the pools of litter that inhibit their growth. This specificity and scope distinguishes the model from competitive Lotka-Volterra and classical plant-soil feedback models. 

Although we have focused on N and litter in grasslands, the model's core structure---resource competition coupled with an inter-specific inhibition mechanism---is quite general.  Interpreted broadly, our model suggests that ecosystems may exhibit bistability and hysteresis in response to changes in resource supply when species not only compete for a limiting resource but also produce inhibitory materials (analogs to litter) such as allelopathic chemicals. Predicted conditions for bistability and hysteresis generalize. For example, the inequalities (\ref{eq:stab-cond}) predict that bistability at a particular resource supply level requires the intrinsic growth advantage of the superior resource competitor to outweigh the detriment of its own inhibitory substance but to be outweighed by detriment of the inferior resource competitor's inhibitory substance. Likewise, equations (\ref{eq:stab-compare}) suggest that hysteresis may result when an increase in the supply of an originally limiting resource allows the intrinsic growth rate of the inferior resource competitor to outpace that of the superior competitor.

\newpage{}
\section*{Appendix: Equilibrium analysis of model (\ref{eq:ode-system})}
\label{sec:appendix}
\renewcommand{\theequation}{B\arabic{equation}}
\renewcommand{\thetable}{B\arabic{table}}
\setcounter{equation}{0}  
\setcounter{figure}{0}
\setcounter{table}{0}

Equilibria of the system (3) can be determined as follows. Imposing the equilibrium conditions $L_n'=L_e'=0$ on equations (3c) and (3d) yields the relationships
\begin{linenomath*}
\begin{equation}
   L_n=\frac{m_n}{k_n}P_n \hspace{1cm} \text{and } \hspace{1cm}  L_e=\frac{m_e}{k_e}P_e,
    \label{eq:equil_L}
\end{equation}
\end{linenomath*}
while the equilibrium condition $P_n'+P_e'+L_n'+L_e'+N'=0$ implies 
\begin{linenomath*}
\begin{equation}
   N=\frac{c+\gamma(d_n+d_e)}{\sigma}. \label{eq:equil_N}
\end{equation}
\end{linenomath*}
\noindent Equations (\ref{eq:equil_L}) and (\ref{eq:equil_N}) allow the elimination of $L_n$, $L_e$, and $N$ from the equations $P_n'=0$ and $P_e'=0$, yielding
\begin{linenomath*}
\begin{align}
   0&=\left(f_n\left(\frac{m_n}{k_n}P_n,\frac{m_e}{k_e}P_e,\frac{c+\gamma(d_n+d_e)}{\sigma}\right)-m_n-u_n\right)P_n+d_n \label{eq:Pndot_subbed}\\
   0&=\left(f_e\left(\frac{m_n}{k_n}P_n,\frac{m_e}{k_e}P_e,\frac{c+\gamma(d_n+d_e)}{\sigma}\right)-m_e-u_e\right)P_e+d_e. \label{eq:Pedot_subbed}
\end{align}
\end{linenomath*}

\noindent Solutions to equations (\ref{eq:Pndot_subbed}) and (\ref{eq:Pedot_subbed}) form two curves in the $P_n,P_e$-plane, whose intersections give the equilibria of the system (see Figure \ref{fig:nullclines}).  It is important to note that these curves are not planar nullclines. Reducing our view to the $P_n$,$P_e$-plane required imposing equilibrium state variable relationships that do not hold during transient dynamics. The resulting planar picture reliably identifies equilibria, not their full five-dimensional stability.

In the simpler case of dispersal ($d_n=d_e=0$), equilibria and some stability conditions can be determined analytically. Let  $r_i(N^*)=\frac{g_i\frac{c}{\sigma}}{h_i+\frac{c}{\sigma}}$.  A trivial equilibrium occurs at $P_n^*=P_e^*=L_n^*=L_e^*=0$, $N^*=c/\sigma$, a native-dominated equilibrium occurs at 
\begin{linenomath*}
\begin{subequations}\label{eq:nEq}
\begin{align}
    P_n^{*,1}&=k_n\ln\left(\frac{r_n(N^*)}{m_n+u_n}\right)/(\beta_{nn}m_n) \label{nEq-Pn} \\
    P_e^{*,1}&=0 \label{nEq-Pe} \\
    L_n^{*,1}&=\ln\left(\frac{r_n(N^*)}{m_n+u_n}\right)/\beta_{nn}\label{nEq-Ln} \\
    L_e^{*,1}&=0 \\
    N^*&=c/\sigma\label{nEq-N} 
\end{align}
\end{subequations}
\end{linenomath*}

\noindent and an exotic dominated equilibrium occurs at 
\begin{linenomath*}
\begin{subequations}\label{eq:eEq}
\begin{align}
    P_n^{*,2}&=0 \label{eEq-Pn} \\
    P_e^{*,2}&=k_e\ln\left(\frac{r_e(N^*)}{m_e+u_e} \right)/(\beta_{ee}m_e) \label{eEq-Pe} \\
    L_n^{*,2}&=0\label{eEq-Ln} \\
    L_e^{*,2}&=\ln\left(\frac{r_e(N^*)}{m_e+u_e} \right)/\beta_{ee}\label{eEq-Le} \\
    N^*&=c/\sigma\label{eEq-N} 
\end{align}
\end{subequations}
\end{linenomath*}

\noindent
A fourth equilibrium is non-negative (yielding coexistence) for certain parameter combinations; its analytic form is unwieldy but it can be found numerically.

Conditions necessary for stability of the native- and exotic-dominated equilibria were determined analytically by calculating the Jacobian for the vector field given by \eqref{eq:ode-system} and evaluating the Jacobian at each equilibrium. The analytic form of these Jacobian matrices are unwieldy, but at each equilibrium the fifth-degree characteristic polynomial has two linear factors that reveal eigenvalues. The native-dominated equilibrium has eigenvalues 
\begin{align*}
    \lambda_1&=r_e(N^*)e^{-\beta_{ne}L_n^{*,1}}-m_e-u_e\\
    \text{and}\hspace{1cm} \lambda_2&=-k_e,
\end{align*}

while the exotic-dominated equilibrium has eigenvalues
\begin{align*}
    \lambda_3&=r_n(N^*)e^{-\beta_{en}L_e^{*,2}}-m_n-u_n\\
    \text{and}\hspace{1cm} \lambda_4&=-k_n.
\end{align*}

\noindent The left inequality in \eqref{eq:stab-cond} was derived from requiring $\lambda_1<0$, substituting $L_n^{*,1}$ from \eqref{nEq-Ln}. The right inequality in  \eqref{eq:stab-cond} was derived from requiring $\lambda_3<0$, substituting $L_e^{*,2}$ from \eqref{eEq-Le}.
Inequalities (3) thus provides necessary but not sufficient conditions for the stability of both native- and exotic- dominated equilibria. 

\newpage{}
\bibliographystyle{plain}
\bibliography{nitrogen2022}

\end{document}